\documentclass{PoS}

\newcommand{\fermi}{\textit{Fermi}/LAT}
\newcommand{\ic}{IC\,310}

\title{The EVN view of the highly variable TeV active galaxy IC\,310}

\ShortTitle{The EVN view of the highly variable TeV active galaxy IC\,310}

\author{\speaker{Robert Schulz}$^{a,b}$, Matthias Kadler$^a$, Eduardo Ros$^{c,d}$, Dorit Eisenacher Glawion$^a$, Uwe Bach$^c$, Dominik Els\"asser$^a$, Christoph Grossberger$^{e,b}$, Ingo Kreykenbohm$^b$, Karl Mannheim$^a$, Cornelia M\"uller$^{a,b}$, Jonas Tr\"ustedt$^a$, J\"orn Wilms$^b$\\
        {$^a$}Lehrstuhl f\"{u}r Astronomie, Universit\"{a}t W\"{u}rzburg, 
        Campus Hubland Nord, Emil-Fischer-Strasse 31, 97074 W\"{u}rzburg, Germany\\
        {$^b$}Dr. Remeis Sternwarte \& ECAP, Universit\"{a}t Erlangen-N\"{u}rnberg, 
        Sternwartstr. 7, 96049 Bamberg, Germany\\
        {$^c$}Max-Planck-Institut f\"{u}r Radioastronomie, 
        Auf dem H\"{u}gel 69, 53121 Bonn, Germany\\
        {$^d$}Observatori Astron\`{o}mic \& Departament d'Astronomia i Astrof\'{\i}sica, Universitat de Val\`{e}ncia, 
        46071 Val\`{e}ncia, Spain\\
        {$^e$}Max-Planck-Institut f\"{u}r extraterrestrische Physik,
        Giessenbachstrasse 1, 85741 Garching, Germany\\
        E-mail: \email{robert.schulz@physik.uni-wuerzburg.de}}

\abstract{Very-high-energy $\gamma$-ray observations of the active galaxy \ic{} with the MAGIC telescopes have revealed fast variability with doubling time scales of less than 4.8\,min. This implies that the emission region in \ic{} is smaller than 20\% of the gravitational radius of the central supermassive black hole with a mass of $3\times 10^8 M_\odot$, which poses serious questions on the emission mechanism and classification of this enigmatic object. We report on the first quasi-simultaneous multi-frequency VLBI observations of \ic{} conducted with the EVN. We find a blazar-like one-sided core-jet structure on parsec scales, constraining the inclination angle to be less than $\sim 20^\circ$ but very small angles are excluded to limit the de-projected length of the large-scale radio jet.}

\FullConference{12th European VLBI Network Symposium and Users Meeting,\\
		7-10 October 2014\\
		Cagliari, Italy}

\begin{document}

\section{Introduction}

The active galaxy \ic{} is located at a distance of $z=0.0189$ \cite{Bernardi2010} in the Perseus cluster and harbours a supermassive black hole with a mass of $3\times 10^8 M_\odot$ \cite{Aleksic2014b}. It was originally classified as a head-tail radio galaxy on kpc-scales \cite{Sijbring1998}. However, Very Long Baseline Array (VLBA) snap-shot observations at 8.3\,GHz revealed a blazar-like morphology on pc-scales with a single-sided jet extending from a dominating core \cite{Kadler2012}. It is part of the MOJAVE monitoring program\footnote{\texttt{http://www.physics.purdue.edu/astro/MOJAVE/index.html}} observing with the VLBA at 15\,GHz.

\ic{} is a source of high-energy emission up to the very-high $\gamma$-ray energies. It was detected above 30\,GeV by \fermi{} \cite{Neronov2010} and at TeV energies by the MAGIC telescopes \cite{Aleksic2010}. In 2009 October and 2010 February MAGIC detected strong flux variability on time scales of one day, concluding that such a short-term variability requires a size of the very high energy emission zone of $\lesssim 80$ Schwarzschild-Radii \cite{Aleksic2014a}. 

A more recent MAGIC observation \cite{Aleksic2014b} in 2012 November revealed an even stronger high-state with multiple flares within a single night. The observation was part of a multiwavelength observation campaign including VLBI observations with the European VLBI Network (EVN). In the resulting first VLBI image of \ic{} at 5.0\,GHz counter-jet emission was not detected, yielding an upper limit of the angle of the jet to the line-of-sight of 20$^\circ$. 
%A lower limit on the angle of the jet was derived to be $\sim$\,5$^\circ$--10$^\circ$ based on geometric constraints on the maximum de-projected size of the large scale radio jet. 
The TeV lightcurve measured by the MAGIC telescopes showed highly rapid flux variations with a doubling time scale down to 4.8\,min. Theoretical considerations to explain the short-term variability lead to a size of the emission zone of 20\% of the gravitational radius of the supermassive black hole, which were interpreted in the framework of particle acceleration models similar to those used for pulsars.

\begin{figure}
	\centering
	\includegraphics[width=.5\textwidth]{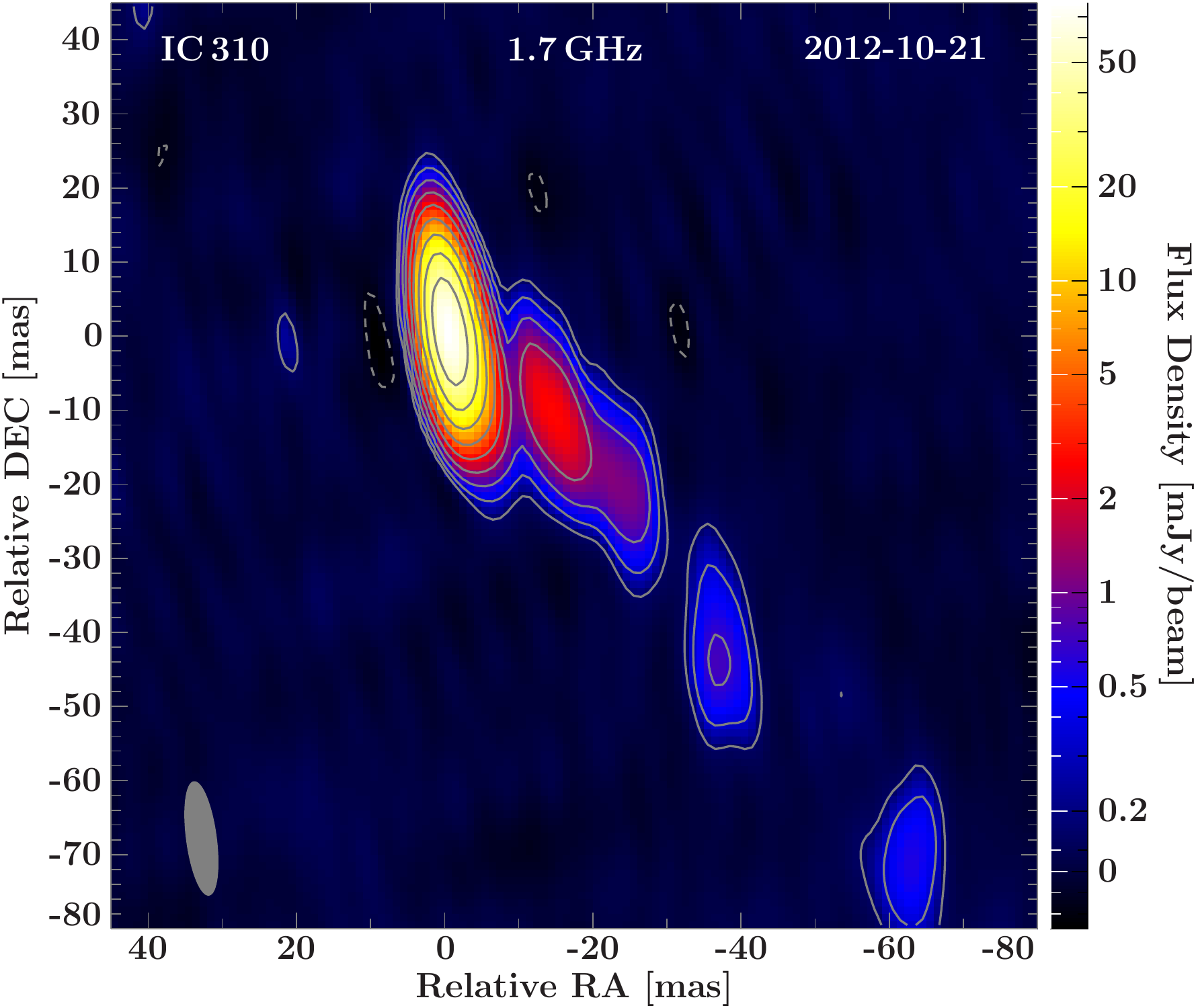}\\[10pt]
	\includegraphics[width=.5\textwidth]{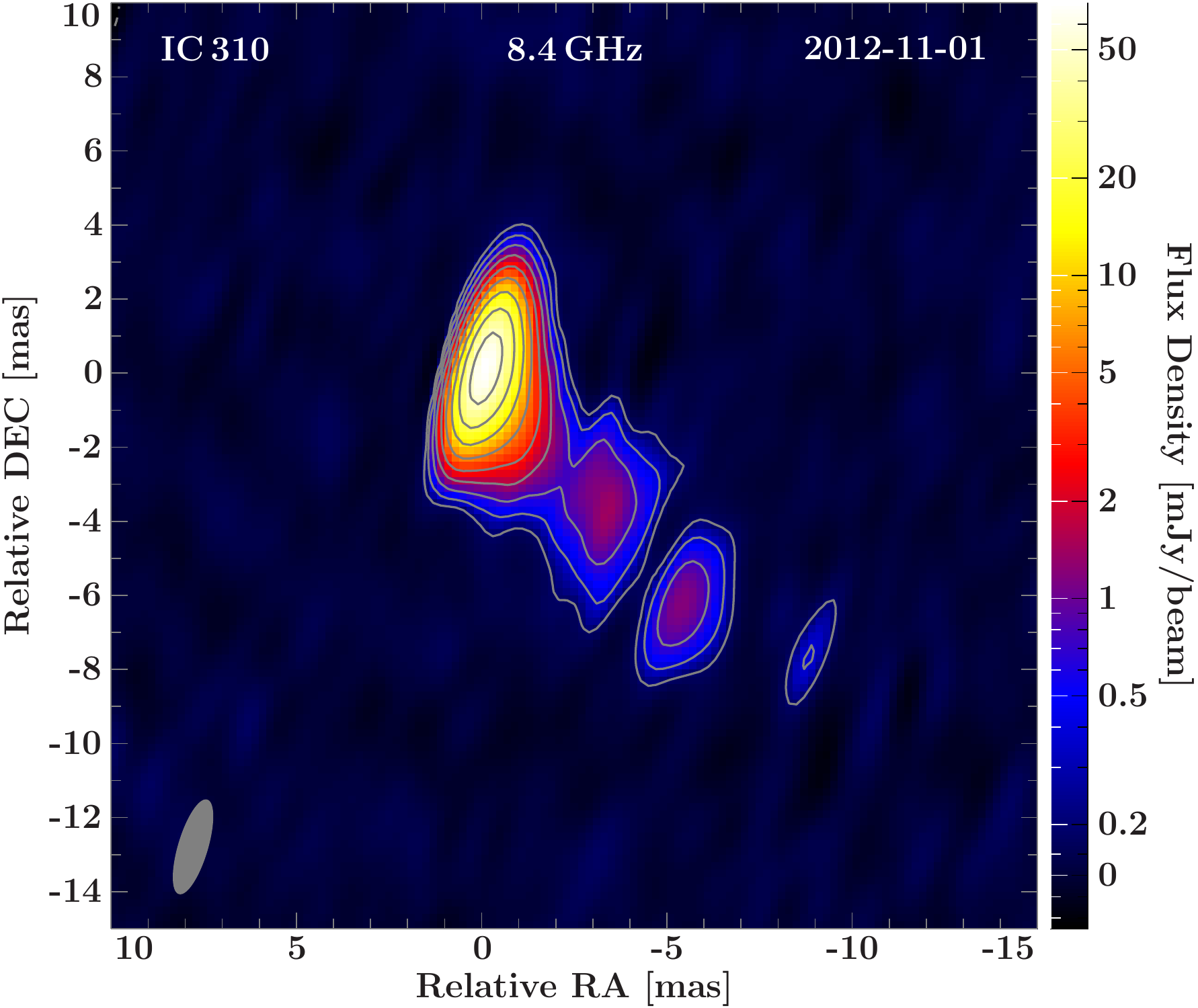}
	\caption{EVN images of \ic{} at 1.7\,GHz (top) and 8.4\,GHz (bottom). The contour lines start at five times the rms noise level $\sigma_\mathrm{rms}$ (see Table 1) and increase logarithmically by factors of two. Negative contours are shown with dashed gray lines. The gray ellipse represents the synthesized beam. 1\,mas correspond to a projected linear scale of 0.39\,pc.}
	\label{fig:EVN}
\end{figure}

\section{The EVN campaign}

The EVN observed \ic{} quasi-simultaneously at 1.7\,GHz, 5.0\,GHz, 8.4\,GHz, and 22\,GHz. The 5.0\,GHz image has been discussed previously \cite{Aleksic2014b}. Here, we present images at 1.7\,GHz and 8.4\,GHz, while results from the 22\,GHz observation will be discussed elsewhere. The observation at 1.7\,GHz yield the first VLBI image of \ic{} at this frequency. The data were calibrated using standard routines in 
\textsc{AIPS} \cite{Greisen2003}. Gain factor corrections had to be determined manually for Jodrell Bank, Zelenchukskaya, and Badary. Hybrid imaging was performed using the software \textsc{DIFMAP} \cite{Shepherd1997}. A list of the participating stations and the image parameters is given in Table \ref{tab:EVN}. 

The resulting new images are shown in Fig. \ref{fig:EVN} and reveal a one-sided jet similar to the 5.0\,GHz observation \cite{Aleksic2014b}. The sensitivity at 8.4\,GHz is improved over the snap-shot VLBA observation at 8.3\,GHz by a factor of $\sim 5$. The total flux density in the EVN images is consistent with a flat spectrum around 0.1\,Jy. The brightness temperature of the prominent core at each frequency (including 5.0\,GHz)  was calculated to be of the order of $\sim10^{10}\mathrm{\,K}$ based on fitting elliptical Gaussian functions to the visibility data.

A comparison of the pc-scale (EVN) and kpc-scale morphology (NVSS, \cite{Condon1998}) does not show any substantial variation of the position angle of the jet on de-projected scales from about $1\mathrm{\,pc}$ out to about $500\mathrm{\,kpc}$, which is consistent with previous studies \cite{Kadler2012}.

\section{Summary \& Outlook}
We present the first EVN images at 1.7\,GHz and 8.4\,GHz of the extremely variable TeV active galaxy \ic{} located in the Perseus cluster. The new VLBI data confirm the remarkably well collimated jet from pc-scales to kpc-scales and reveal a flat radio spectrum on parsec scales. The one-sided jet morphology extends from a high brightness temperature VLBI core. A detailed analysis of the full EVN campaign in particular in terms of the spectral index distribution and in combination with single-dish measurements will be featured in an upcoming publication (Schulz et al., in preparation).

\acknowledgments
{\small R.S., C.M. and J.T. wish to thank RadioNet3 for support to attend this meeting. E.R. was partially supported by the Spanish MINECO project AYA2012-38491-C02-01 and by the Generalitat Valenciana project PROMETEOII/2014/057. We acknowledge support by the COST MP0905 action `Black Holes in a Violent Universe'. The research leading to these results has received funding from the European Commission Seventh Framework Programme (FP/2007-2013) under grant agreement No. 283393 (RadioNet3). The European VLBI Network is a joint facility of European, Chinese, South African and other radio astronomy institutes funded by their national research councils. This research has made use of the Interactive Spectral Interpretation System (ISIS). This research has made use of a collection of ISIS scripts provided by the Dr. Karl Remeis observatory, Bamberg, Germany at \texttt{\footnotesize http://www.sternwarte.uni-erlangen.de/isis/}.
}

	\begin{table}
		{\centering
			\caption[]{Overview of observation and image parameters}
			\begin{tabular}{@{}c@{\,\,\,}c@{\,\,\,}p{4.2cm}@{\,\,\,}c@{\,\,\,}c@{\,\,\,}c@{\,\,\,}c@{}}
				\hline
				$\nu$ & Date & Array$^a$ & $S_\mathrm{tot}^b$ & $S_\mathrm{peak}^c$ & $\sigma_\mathrm{rms}^d$ & Beam$^e$\\
				$[$GHz$]$ &   [yyyy-mm-dd]  &  & [Jy] & [Jy/Beam] & [mJy/Beam] & [mas$\times$mas,deg]	 \\
				\hline
				\hline
				1.7 & 2012-10-21 & EF\,WB\,JB\,ON\,MC\,NT\,TR ZC\,BD\,UR\,SH & 0.10 & 0.078 & 0.033 & 15.4$\times$4.0,7.6\\
				8.4 & 2012-11-01 & EF\,WB\,ON\,MC\,YS\,SV\,ZC BD\,UR & 0.10 & 0.070 & 0.037 & 2.6$\times$0.78,$-$16.7	\\
				\hline
			\end{tabular}
			\label{tab:EVN}}
		{\footnotesize 
			{}$^a$ Antenna	EF: Effelsberg, WB: Westerbork, JB: Jodrell Bank, ON: Onsala, MC: Medicina, NT: Noto, TR: Torun, ZC: Zelenchukskaya, BD: Badary, UR: Urumqi, SH: Sheshan, YS: Yebes, SV Svetloe; 
			{}$^b$ total flux density;
			{}$^c$ peak flux density;
			{}$^d$ noise level;
			{}$^e$ beam parameters (major axis $\times$ minor axis, position angle)
		}
	\end{table}

\end{document}